\documentclass[arxiv,usenatbib]{iupartex}
\usepackage{newtxtext,newtxmath}
\usepackage[T1]{fontenc}
\usepackage{ae,aecompl}

\usepackage{multicol}   
\usepackage{bm}		    
\usepackage{pdflscape}	

\title[Collinder 74]{Comprehensive Analysis of the Open Cluster Collinder 74}

\author[Yontan \& Canbay]{%
T. Yontan$^{1\cc}$\orcid{0000-0002-5657-6194},
and
R. Canbay$^{2}$ \orcid{0000-0003-2575-9892}
\affsep \\
$^1$Istanbul University, Faculty of Science, Department of Astronomy and Space Sciences, 34119, Beyaz\i t, Istanbul, Turkey\\
$^2$Istanbul University, Institute of Graduate Studies in Science, Programme of Astronomy and Space Sciences, 34116, Beyaz{\i}t, Istanbul, Turkey\\
}

\corres{T. Yontan}{talar.yontan@istanbul.edu.tr}

\pubyear{2023}

\doiheader{XXXXXXX/PAR.20XX.00000}
\date{
	\pSubmit{XX.XX.XXXX} 
	\pRevReq{XX.XX.XXXX}
	\pLastRevRec{XX.XX.XXXX}
	\pAccept{XX.XX.XXXX}
	\pPubOnl{XX.XX.XXXX0}
}
\volume{0}
\volnumber{1}

\begin{document}
\label{firstpage}
\pagerange{\pageref*{firstpage}--\pageref*{lastpage}}
\maketitle

\begin{abstract}
In this study, we have used the {\it Gaia} Third Data Release (Gaia DR3) to investigate an intermediate-age open cluster Collinder 74. Taking into account the stars with membership probabilities over 0.5 and inside the limiting radius of the cluster, we identified 102 most likely cluster members. The mean proper-motion components of Collinder 74 are estimated as ($\mu_{\alpha}\cos \delta, \mu_{\delta})=(0.960 \pm 0.005, -1.526 \pm 0.004$) mas yr$^{-1}$. We detected previously confirmed four blue straggler stars which show flat radial distribution. Colour excess, distance, and age of the cluster were estimated simultaneously by fitting {\sc PARSEC} isochrones to the observational data on {\it Gaia} based colour magnitude diagram. These values were derived as $E(G_{\rm BP}-G_{\rm RP})=0.425\pm 0.046$ mag, $d=2831 \pm118$ pc and $t=1800 \pm 200$ Myr, respectively. The mass function slope was estimated as $\Gamma = 1.34 \pm 0.21$ within the mass range $0.65\leq M/ M_{\odot}\leq 1.58$ which is well matched with that of Salpeter. Stellar mass distribution indicated that the massive and most likely stars are concentrated around the cluster center. The total mass of the cluster was found to be 365 $M/ M_{\odot}$ for the stars with probabilities $P>0$. Galactic orbit integration shows that the Collinder 74 follows a boxy pattern outside the solar circle and is a member of the thin-disc component of the Galaxy. 
\end{abstract}




\section{Introduction}
\label{sec:Introduction}

Open star clusters (OCs), also known as Galactic clusters, are loosely bound groups of stars that emerged from the same molecular cloud, sharing a common origin and age. As relatively young and dynamically active systems, OCs typically contain hundreds to thousands of stars that share similar chemical composition, age, and distance. The formation origin of OCs plays a crucial role in our understanding of stellar formation and evolution and makes them ideal laboratories for studying stellar properties, such as temperature, luminosity, and mass \citep{Lada_2003, Kim_2017}. Also, member stars of OCs have similar movement directions in the sky which knowledge makes proper-motion components useful tools to separate physical members from the field star contamination \citep{Sariya_2021}. This method provides a reliable sample of member stars for the estimation of astrophysical parameters for OCs \citep{Bisht_2020, Sariya_2021}.  

In this study, we estimated the structural, kinematic, and astrophysical properties of Collinder 74 (Coll 74) open cluster. The cluster is positioned at $\alpha=05^{\rm h} 48^{\rm m} 40^{\rm s}.8, \delta= +07^{\circ} 22^{\rm '} 26^{\rm''}.4$ (J2000) corresponding to Galactic coordinates $l=199^{\circ}.019, b=-10^{\circ}.379$ according to \citet{Cantat-Gaudin_2020}. Coll 74 is a centrally concentrated old open cluster located in the third Galactic quadrant toward the Galactic anti-center region. \citet{Ann_1999} afforded {\it UBVI} CCD photometry and suggested that the age of the cluster is $1300 \pm 200$ Myr and it is located at $2511 \pm 245$ pc from the Sun. \citet{Tadross_2001} performed {\it UBV} CCD observations and defined colour excess, distance and age of the cluster as $E(B-V)=0.38$ mag, $d=2254$ pc and $t=1600$ Myr, respectively. \citet{Dias_2006} presented the kinematics of the cluster using UCAC2 Catalogue positions and proper motions. They derived proper-motion components as ($\mu_{\alpha}\cos\delta, \mu_{\delta})=(-0.49, -3.49)$ mas yr$^{-1}$. \citet{Carraro_2007} investigated CCD photometry in the $V$ and $I$-bands and obtained colour excess, distance and age of the cluster as $E(B-V)=0.28\pm0.04$ mag, $d=1500$ pc and $t=3000$ Myr, respectively. From the analyses of $BVI$ photometry \citet{Hasegawa_2008} concluded that the Coll 74 is 3680 pc distant from the Sun and 1400 Myr old cluster. \citet{Dias_2014} derived the mean proper-motion components by taking into account the U.S. Naval Observatory CCD Astrograph Catalogue \citep[UCAC4;][]{Zacharias_2013} as ($\mu_{\alpha}\cos\delta, \mu_{\delta})=(1.92 \pm 0.55, -3.00 \pm 0.10$) mas yr$^{-1}$. \citet{Loktin_2017} redetermined main parameters using published photometric measurements provided by 2MASS catalogue for 959 clusters including Coll 74. They estimated colour excess, distance and age as $E(B-V)=0.418$ mag, $d=3293$ pc and $t=1400$ Myr, respectively. Also, from RAVE catalogue \citet{Loktin_2017} estimated proper-motion components in equatorial coordinate system as ($\mu_{\alpha}\cos\delta, \mu_{\delta})= (0.294 \pm 0.181, -1.091 \pm 0.172$) mas yr$^{-1}$, respectively.

\begin{table*}
{\normalsize
\setlength{\tabcolsep}{7pt}
\renewcommand{\arraystretch}{1.2}
\small
  \centering
  \caption{Basic parameters of the Collinder 74 open cluster collected from the literature.}
  \begin{tabular}{ccccccccc}
    \hline
    \hline
$E(B-V)$ & $\mu_{\rm V}$ & $d$ & [Fe/H] & $t$ &  $\langle\mu_{\alpha}\cos\delta\rangle$ &  $\langle\mu_{\delta}\rangle$ & $V_{\gamma}$ & Ref \\
(mag) & (mag) & (pc)  & (dex) & (Myr) & (mas yr$^{-1}$) & (mas yr$^{-1}$) & (km s$^{-1})$ &      \\
\hline
0.38$\pm$0.04             & 13.08$\pm$0.25   & 2511$\pm$245         & 0.07             & 1300$\pm$200            & $-$              & $-$              & $-$             & (01) \\
0.38                      & 13.00            & 2254                 & 0.054            & 1600                    & $-$              & $-$              & $-$             & (02) \\
0.38                      & 13.21            & 2549                 & $-$              & 1300                    & $-$              & $-$              & $-$             & (03) \\
$-$                       & $-$              & $-$                  & $-$              & $-$                     & -0.49            & -3.49            & $-$             & (04) \\
0.28$\pm$0.04             & 11.75$\pm$0.10   & 1500                 & $-$              & 3000                    & $-$              & $-$              & $-$             & (05) \\  
0.36                      & 13.95            & 3680                 & -0.38            & 1400                    & $-$              & $-$              & $-$             & (06) \\
0.604                     & 13.98            & 2637                 & $-$              & 1230                    & 0.91             & -3.86            & $-$             & (07) \\
$-$                       & $-$              & $-$                  & $-$              & $-$                     & 1.92$\pm$0.55    & -3.00$\pm$0.10   & $-$             & (08) \\
$-$                       & $-$              & 2510                 & 0.050            & 1288                    & $-$              & $-$              & $-$             & (09) \\
0.418                     & 12.588           & 3293                 & $-$              & 1400                    & 0.294$\pm$0.181  & -1.091$\pm$0.172 & $-$             & (10) \\
$-$                       & $-$              & $-$                  & $-$              & $-$                     & 0.77$\pm$2.15    & 0.47$\pm$3.54    & $-$             & (11) \\
$-$                       & $-$              & $2453_{-484}^{+797}$ & $-$              & $-$                     & 1.011$\pm$0.016  & -1.512$\pm$0.016 & $-$             & (12) \\
$-$                       & $-$              & $2453_{-484}^{+797}$ & $-$              & $-$                     & 1.011$\pm$0.016  & -1.512$\pm$0.016 & 20.18$\pm$0.39  & (13) \\
$-$                       & $-$              & 2747$\pm$332         & $-$              & 2190$\pm$131            & 0.981$\pm$0.200  & -1.497$\pm$0.199 & $-$             & (14) \\
$-$                       & $-$              & $2453_{-484}^{+797}$ & -0.05$\pm$0.03   & $-$                     & 1.011$\pm$0.121  & -1.512$\pm$0.122 & 15.94$\pm$17.57 & (15) \\
0.274                     & 11.99            & 2498$\pm$494         & $-$              & 1900                    & 1.011$\pm$0.121  & -1.512$\pm$0.122 & $-$             & (16) \\
0.391$\pm$0.076           & $-$              & 2153$\pm$144         & -0.083$\pm$0.084 & 2760                    & 0.995$\pm$0.170  & -1.528$\pm$0.175 & 20.20$\pm$0.80  & (17) \\ 
$-$                       & $-$              & 2356                 & $-$              & 2100                    & 1.011$\pm$0.121  & -1.512$\pm$0.122 & 20.18$\pm$0.39  & (18) \\ 
0.511$\pm$0.074           & $-$              & 2466$\pm$22           & $-$              & 627$\pm$348             & 0.964$\pm$0.007  & -1.546$\pm$0.007 & 20.93$\pm$4.10  & (19) \\ 
0.301$\pm$0.033           & 13.052$\pm$0.088 & 2831$\pm$118         & 0.052$\pm$ 0.034 & 1800$\pm$200            & 0.960$\pm$0.005  & -1.526$\pm$0.004 & 20.55$\pm$0.41  & (20) \\ 

  \hline
    \end{tabular}%
    \\
(01) \citet{Ann_1999}, (02) \citet{Tadross_2001}, (03) \citet{Lada_2002}, (04) \citet{Dias_2006}, (05) \citet{Carraro_2007}, (06) \citet{Hasegawa_2008}, (07) \citet{Kharchenko_2013}, (08) \citet{Dias_2014},  (09) \citet{Marsakov_2016}, (10) \citet{Loktin_2017}, (11) \citet{Dias_2018}, (12) \citet{Cantat-Gaudin_2018}, (13) \citet{Soubiran_2018}, (14) \citet{Liu_2019}, (15) \citet{Zhong_2020}, (16) \citet{Cantat-Gaudin_2020}, (17) \citet{Dias_2021}, (18) \citet{Tarricq_2021}, (19) \citet{Hunt_2023}, (20) This study
  \label{tab:literature}%
}
  \end{table*}%

With the first data release of the {\it Gaia} \citep{Gaia_DR1},  many researchers investigated the astrometric and kinematic properties of the Coll 74. According to literature studies performed with {\it Gaia} data, values of the mean radial velocity of the cluster differs from $15.94 \pm 17.57$ km s$^{-1}$ \citep{Zhong_2020} to $20.20 \pm 0.80$ km s$^{-1}$ \citep{Dias_2021} and distance to the Sun changes between 1500 pc \citep{Carraro_2007} and 3680 pc \citep{Hasegawa_2008}. Also, the age of the cluster varies from 1230 Myr \citep{Kharchenko_2013} to 3000 Myr \citep{Carraro_2007}. The literature results are listed in Table \ref{tab:literature} for detailed comparison. The main purpose of this study is to find out structural, astrophysical, and kinematic properties of the Coll 74 open cluster.
 
\section{Data}

\begin{figure}
\centering
\includegraphics[scale=0.70, angle=0]{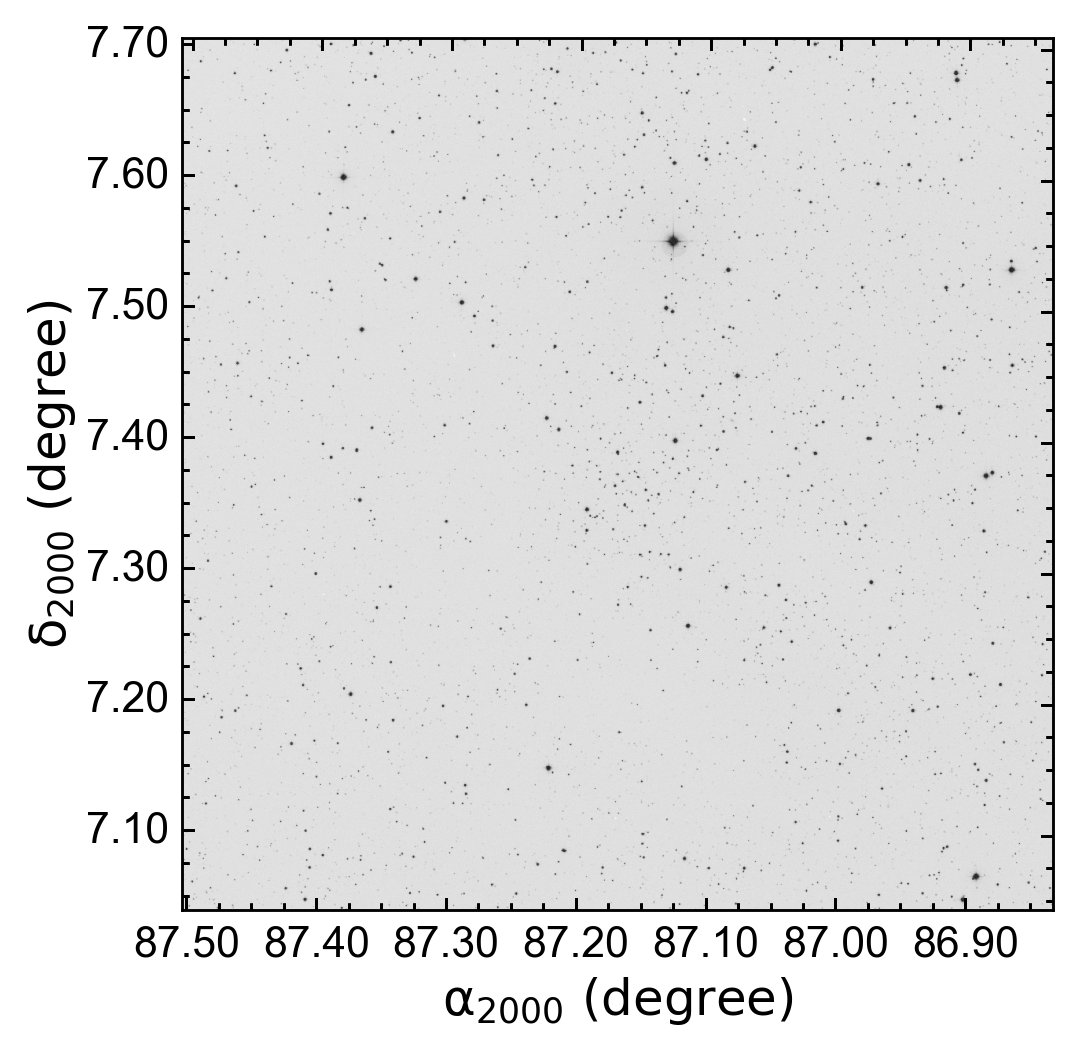}
\caption{ Finding chart of the Coll 74 for $35'\times 35'$ region. Up and left directions represent North and East, respectively.} 
\label{fig:ID_charts}
\end {figure}

The astrometric, photometric, and spectroscopic data for Coll 74 open cluster was taken from {\it Gaia}'s third data release \citep[{\it Gaia} DR3,][]{Gaia_DR3}. To do this, we used the central equatorial coordinates of \citet{Cantat-Gaudin_2020} $\langle\alpha, \delta\rangle = (05^{\rm h} 48^{\rm m} 40^{\rm s}.8, +07^{\circ} 22^{\rm '} 26^{\rm''}.4$) and gathered the detected stars in the direction of the cluster for 35$^{\rm '}$-radius field. Hence, we reached 73,326 stars within the applied radius. The finding chart of the Coll 74 ($35' \times 35'$) is shown in Figure~\ref{fig:ID_charts}. The main cluster catalogue contains each stars' position ($\alpha$, $\delta$), photometric magnitude and colour index ($G$, $ G_{\rm BP}-G_{\rm RP}$), trigonometric parallax ($\varpi$), proper-motion components ($\mu_{\alpha}\cos\delta$, $\mu_{\delta}$), radial velocity ($V_\gamma$) and their errors within the $8<G\leq22$ mag.   

\begin{figure}
\centering
\includegraphics[scale=0.8, angle=0]{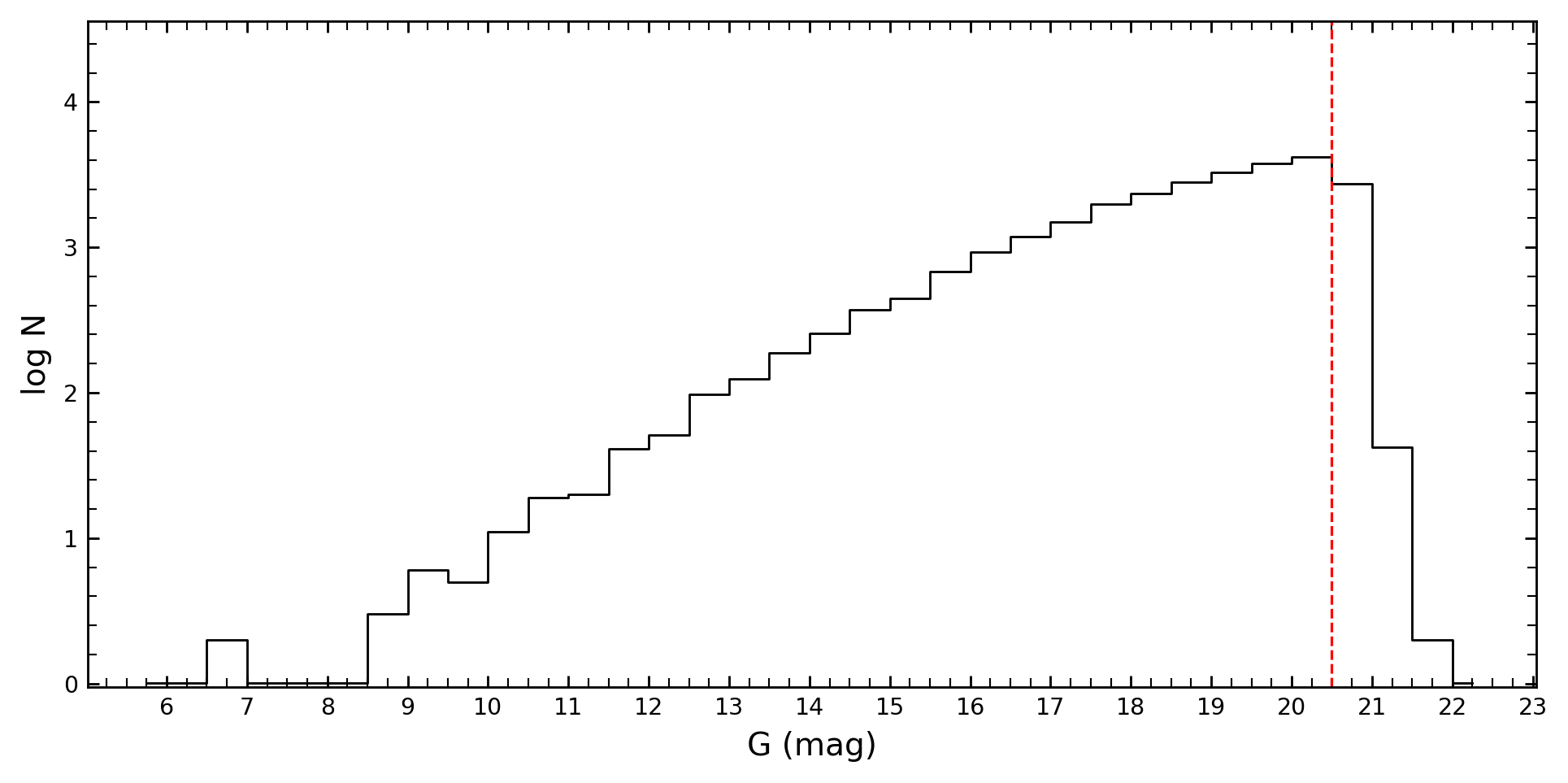}\\
\caption{Distribution of the stars in the direction of Coll 74 for $G$ magnitude intervals. The photometric completeness limit is indicated by a red dashed line.}
\label{fig:histograms}
\end {figure}

\begin{figure}
\centering
\includegraphics[scale=1.2, angle=0]{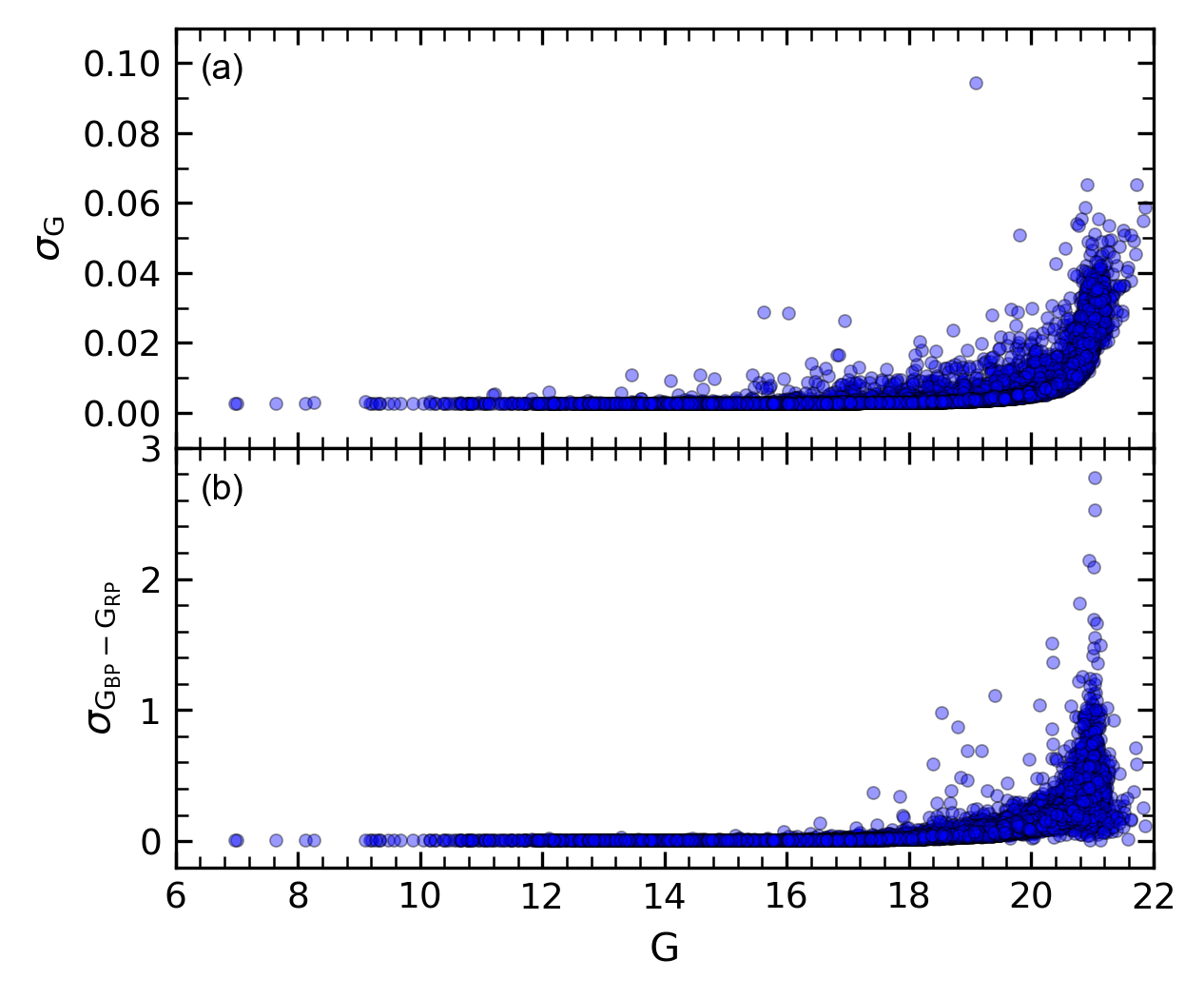}
\caption{~Distribution of mean photometric errors obtained for $G$ apparent magnitude (a) and $G_{\rm BP}-G_{\rm RP}$ colour index versus $G$ magnitude intervals.} 
\label{fig:photometric_errors}
\end {figure}

To achieve reliable structural and astrophysical parameters for Coll 74, we obtained a faint limited magnitude of the used data. For this, we calculated the number of stars that correspond to the $G$ magnitude intervals. The histogram of a number of stars versus $G$ magnitudes is shown in Figure~\ref{fig:histograms}, where a number of stars rise towards the fainter $G$ magnitudes and declines after a certain limit. This limit value is $G=20.5$ mag for the Coll 74 and in the following analyses, we used only the stars brighter than $G=20.5$ mag. We calculated the mean photometric errors of the stars for $G$ magnitude intervals. The mean errors for $G$ and $G_{\rm BP}-G_{\rm RP}$ colour indices reach up 0.011 and 0.228 mag for $G=20.5$ limiting magnitude, respectively. The photometric errors for $G$ magnitudes and $G_{\rm BP}-G_{\rm RP}$ colour indices versus $G$ magnitude intervals are shown in Figure \ref{fig:photometric_errors}.

\section{Results}
\subsection{Spatial structure of Collinder 74}
\begin{figure}[]
\centering
\includegraphics[scale=0.4, angle=0]{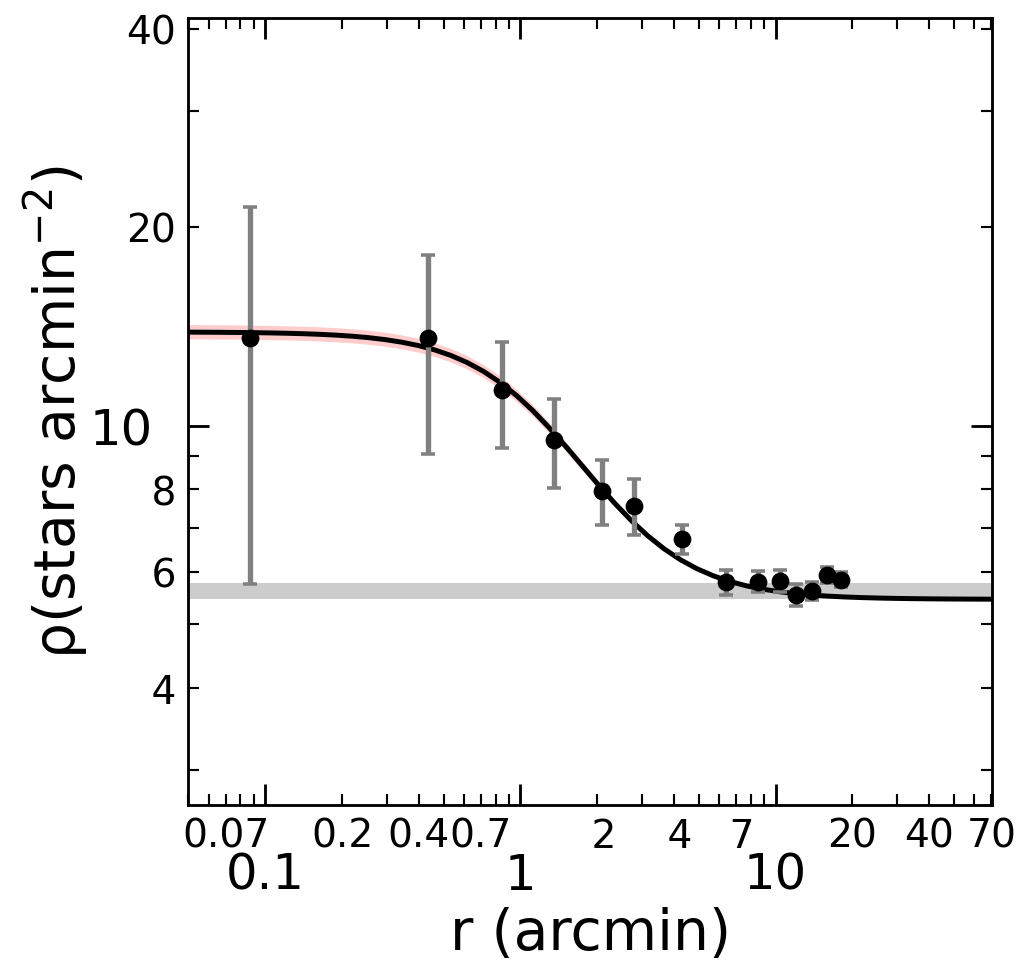}\\
\caption{The RDP of \citet{King_1962} for Coll 74. Stellar density errors were determined from Poisson statistics $1/\sqrt N$, where $N$ is the number of stars. The fitted black curve and horizontal grey shaded area show the best-fitted RDP and background stellar density, respectively.  Also, red-shaded area indicates the $1\sigma$ uncertainty of the fit.} 
\label{fig:king}
\end {figure}

To interpret stellar distribution within the cluster we constructed the radial density profile (RDP) considering adopted central equatorial coordinates presented by \citet{Cantat-Gaudin_2020}. We divided the $35'$ cluster region into many concentric rings surrounding the cluster center and calculated stellar densities ($\rho(r)$) from the stars within $G \leq 20.5$ mag. The stellar densities in $\it i^{\rm th}$ ring were calculated by the equation of $R_{i}=N_{i}/A_{i}$, where, $N_{i}$ and $A_{i}$ indicate the number of stars and area of related ring, respectively. To visualise RDP, we plotted stellar density distribution versus distance from cluster center and fitted the empirical \citet{King_1962} model identified by the following equation: 

\begin{equation}
\rho(r)=f_{\rm bg}+\frac{f_0}{1+(r/r_{\rm c})^2}
\end{equation}
Here, $r$ is the radius of the cluster. The $f_{\rm bg}$, $f_0$, $r_{\rm c}$ are the background stellar density, the central stellar density and the core radius, respectively. We used the $\chi^{2}$ minimisation technique for RDP analyses and estimated $f_{\rm bg}$, $f_0$ and $r_{\rm c}$. In Figure \ref{fig:king} we showed the best-fit result of RDP which is demonstrated by the black solid line. It can be seen from the figure that stellar density is higher near the cluster center and it flattens toward the outer region of the cluster and at a point merges with the field star density. This point is described as limiting radius ($r_{\rm lim}$) and visually adopted as $10'$. The best-fit solution of RDP analyses resulted that the structural parameters to be $f_0= 8.42\pm0.35$ stars arcmin$^{-2}$, $f_{\rm bg}=5.45\pm0.16$ stars arcmin$^{-2}$ and $r_{\rm c}=1.38\pm0.12$ arcmin for the Coll 74.

\subsection{Membership Analyses of Collinder 74}

Working with a sample of stars that are part of the cluster itself is crucial to accurately characterise its properties, such as its age, mass, luminosity function, and dynamics. Field star contamination may contribute to noise and bias during the analyses and can cause inaccurate results. Therefore, field star separation is necessary to understand the astrophysical, astrometric, and kinematic properties of the studied cluster. As their physical formation processes, cluster members are gravitationally bound to the cluster. Therefore, they exhibit similar vectorial movements across the sky relative to the background field stars. Through the analysis of proper-motion components, stars that share the motion of the cluster can be identified as cluster members. Hence, proper-motion components are useful tools to separate cluster members and calculate their membership probabilities \citep{Sariya_2021, Bisht_2020}. The precise astrometric data from {\it Gaia} DR3 data \citep{Gaia_DR3} provide a crucial role for membership analyses.

We used the method of Photometric Membership Assignment in Stellar Clusters \citep[{\sc UPMASK},][]{Krone-Martins_2014} considering {\it Gaia} DR3 astrometric data to calculate membership probabilities of stars in the region of the Coll 74. {\sc UPMASK} uses a clustering algorithm to group stars that have similar positions, proper-motion components, trigonometric parallaxes and are close to each other in space. The algorithm then assigns membership probabilities to each star based on its likelihood of belonging to a particular cluster. We utilised the method in five-dimensional astrometric space considering astrometric measurements ($\alpha$, $\delta$, $\mu_{\alpha}\cos \delta$, $\mu_{\delta}$, $\varpi$), also their uncertainties, of each star. To determine stars' membership probabilities ($P$) we run the program with 100 iterations. As a result, taking into account the stars within both the estimated limiting radius ($r_{\rm lim}$=10$^{\rm '}$) and the completeness limit $G=20.5$ mag, for the open cluster Coll 74 we obtained 102 most likely member stars with membership probabilities of $P\geq 0.5$. These stars are used in the estimation of astrometric and astrophysical parameters of Coll 74.

\begin{figure}
\centering
\includegraphics[scale=0.70, angle=0]{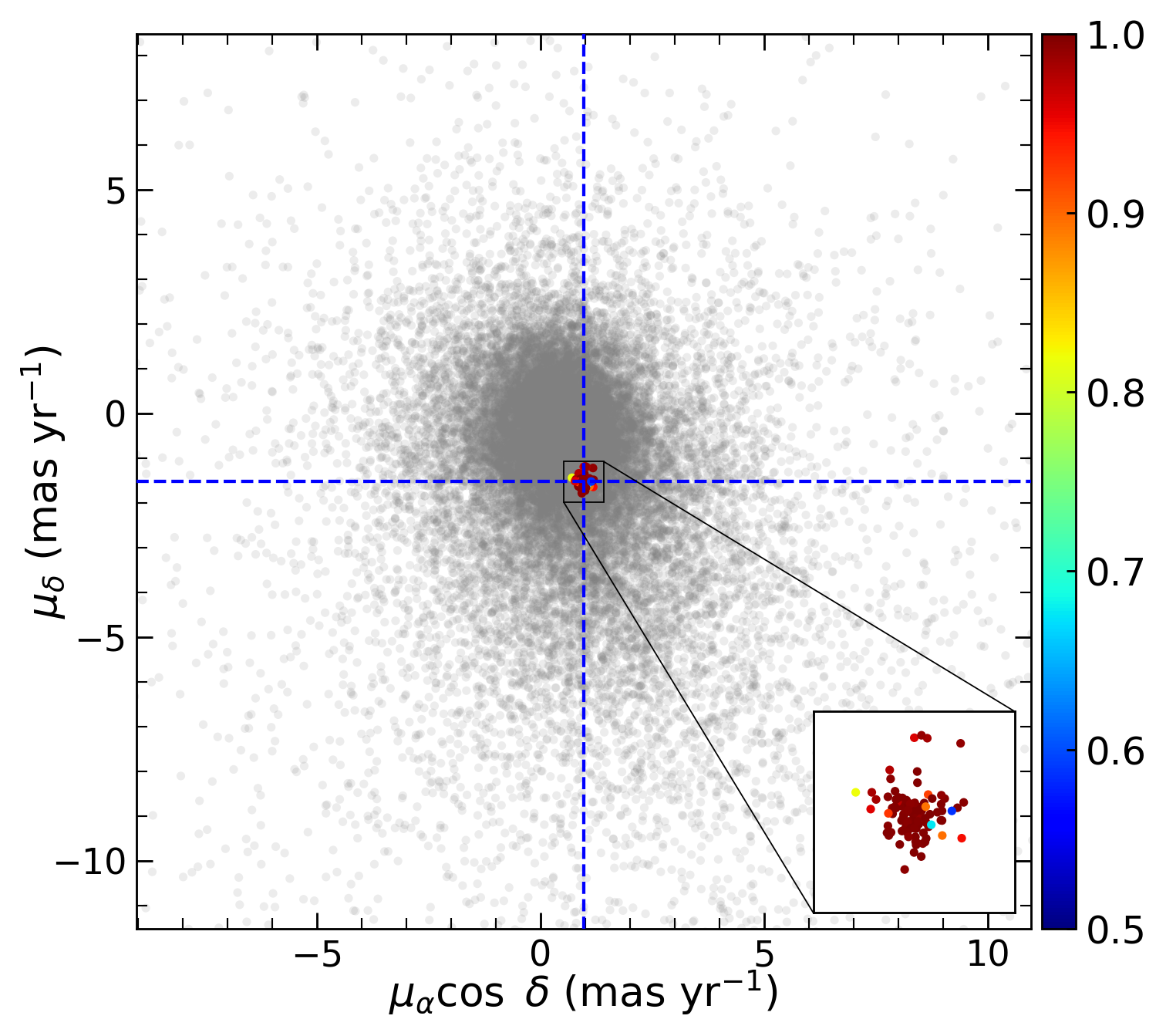}\\
\caption{VPD of Coll 74. The colour scale on the right panel indicates the membership probabilities of the stars for the cluster. The enlarged panel inset shows the cluster's denser region in VPD. The intersection of the dashed blue lines represents the mean proper-motion value for Coll 74.
\label{fig:VPD_all}} 
 \end {figure}

\begin{figure}
\centering
\includegraphics[scale=.70, angle=0]{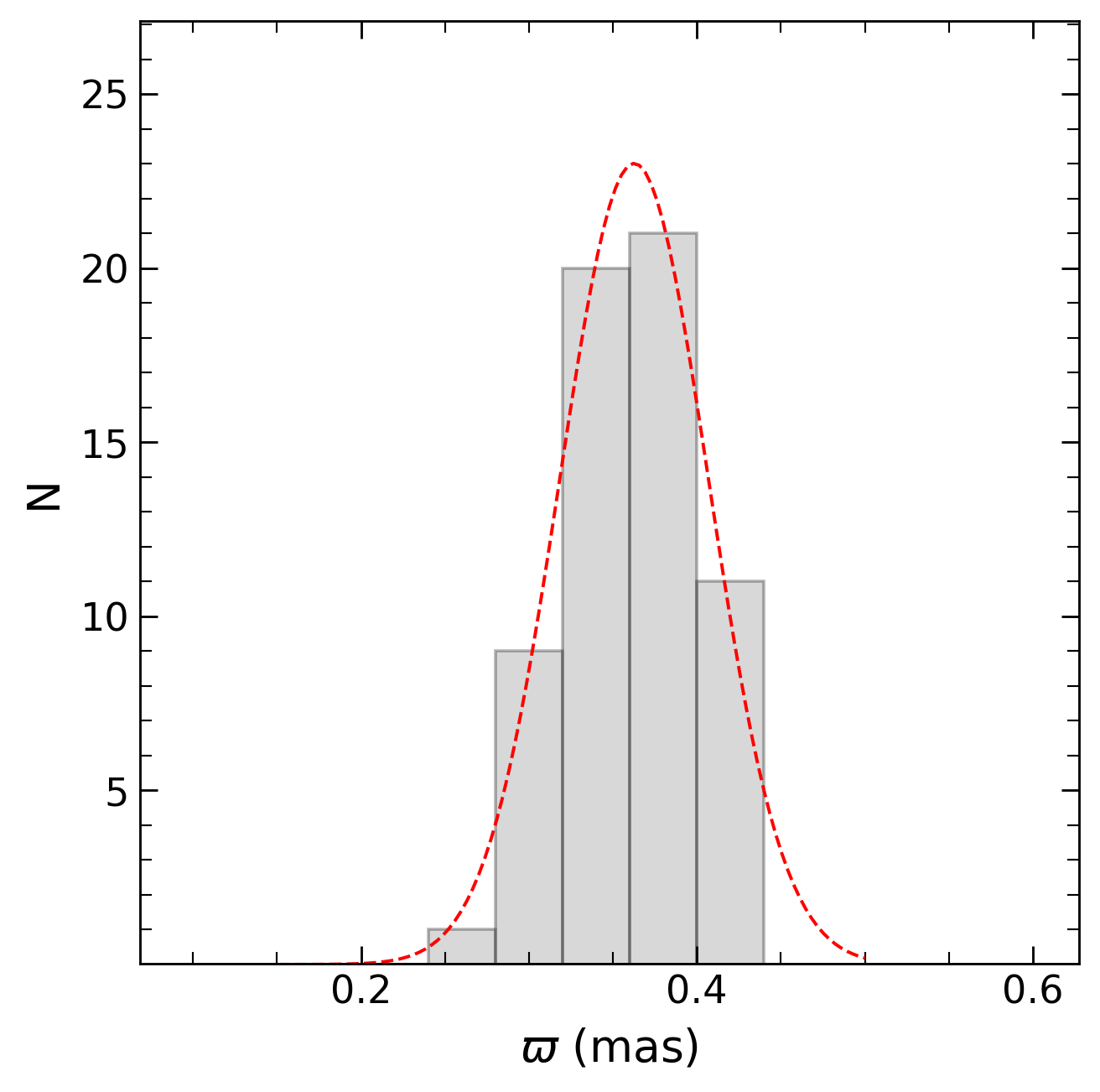}\\
\caption{Histogram of star count of the most likely members ($P\geq 0.5$) in trigonometric parallaxes. The red dashed line indicates the fitted Gaussian function.
\label{fig:plx_hist}}
\end {figure}

To visualise the clustering of the most likely member stars, we plotted vector-point diagram (VPD) by using proper-motion components of the stars and showed it in Figure ~\ref{fig:VPD_all}. It is evident from the VPD that Coll 74 is embedded in the field stars. Even if this is the case, the cluster structure can be distinguished by investigation of the probability values of the stars. We estimated mean proper-motion values from the stars with membership probabilities greater than 0.5 and found the values as ($\mu_{\alpha}\cos \delta, \mu_{\delta})=(0.960 \pm 0.005, -1.526 \pm 0.004$) mas yr$^{-1}$. The trigonometric parallax histogram of the most likely member stars is shown in Figure~\ref{fig:plx_hist}. By fitting the Gaussian function to the histogram, we estimated the mean trigonometric parallax as $\varpi=0.363\pm0.043$ mas for the Coll 74 and corresponding distance value (with the linear equation $d({\rm pc})=1000/\varpi$) mas as $d_{\varpi}=2755\pm326$ pc. This value of distance is close to the values estimated in {\it Gaia} era, as listed in Table~\ref{tab:literature}.    

\subsection{The Blue Straggler Stars of Collinder 74}

Blue straggler stars (BSSs) in open clusters defy the natural aging process by appearing younger and bluer than their surrounding companions. While most stars in open clusters follow common evolutionary processes, BSSs defy the laws of stellar evolution in the cluster by reversing this trend. Interactions between stars in binary systems or stellar collisions within the dense cluster environment are among the leading formation mechanisms for BSSs \citep{Sandage_1953, Zinn_1976, Hills_1976}. 

We identified four BSSs in Coll 74. These stars are confirmed as cluster members with membership probabilities $P\geq0.9$. The positions of BSSs are shown in Figure ~\ref{fig:figure_age}. \citet{Rain_2021} defined five BSSs by using {\it Gaia} DR2 \citep{Gaia_DR2} photometric and astrometric data. Due to the membership analyses being based on {\it Gaia} DR3 data and we took into account the stars within limiting radius ($r_{\rm lim}\leq 10'$), one of the \citet{Rain_2021} BSS was not considered. \citet{Jadhav_2021} using {\it Gaia} DR2 data and identified BSSs in 1246 
 open clusters. They found four BSSs members in Coll 74 which are the same star sample as presented in this study.

\citet{Ferraro_2012} considering the radial distribution of BSSs in the clusters, defined three classes for BSSs. Three of BSSs in Coll 74 are located at a radial distance 0.42, 0.88, and 0.98 arcmin, whereas one star is located at 6.25 arcmin. According to their radial distribution and the criterion of \citet{Ferraro_2012}, we can conclude that the BSSs of Coll 74 show flat distribution and cluster belongs to the family I. The formation mechanisms of blue stragglers in family I clusters are thought to be dominated by stellar collisions and mass transfer in close binary systems. 

\subsection{Astrophysical Parameters of Collinder 74}

To derive age, distance modulus and colour excess of Coll 74, we fitted theoretical {\sc PARSEC} isochrones of \citet{Bressan_2012} to the observed CMD constructed from the most likely cluster members. The age, distance modulus and colour excess of the cluster were estimated simultaneously, while the metallicity of Coll 74 was taken from \citet{Zhong_2020} as [Fe/H]=$-0.052\pm 0.034$ dex. We fitted {\sc PARSEC} models by taking into account the morphology of the cluster and reached the best fit. For the selection of the isochrones and estimation of astrophysical parameters, we transformed the assumed metallicity ([Fe/H]=$-0.052\pm 0.034$ dex) to the mass fraction $z$. To do this, we applied the equation of Bovy\footnote{https://github.com/jobovy/isodist/blob/master/isodist/Isochrone.py} that are available for {\sc PARSEC} models \citep{Bressan_2012}. The equations are given as follow:

\begin{equation}
z_{\rm x}={10^{{\rm [Fe/H]}+\log \left(\frac{z_{\odot}}{1-0.248-2.78\times z_{\odot}}\right)}}
\end{equation}      
and
\begin{equation}
z=\frac{(z_{\rm x}-0.2485\times z_{\rm x})}{(2.78\times z_{\rm x}+1)}.
\end{equation} 
where $z_{\rm x}$ and $z_{\odot}$ are intermediate values where solar metallicity $z_{\odot}$ was adopted as 0.0152 \citep{Bressan_2012}. The calculated mass fraction is $z=0.0136$ for Coll 74.

\begin{figure}
\centering
\includegraphics[scale=0.9, angle=0]{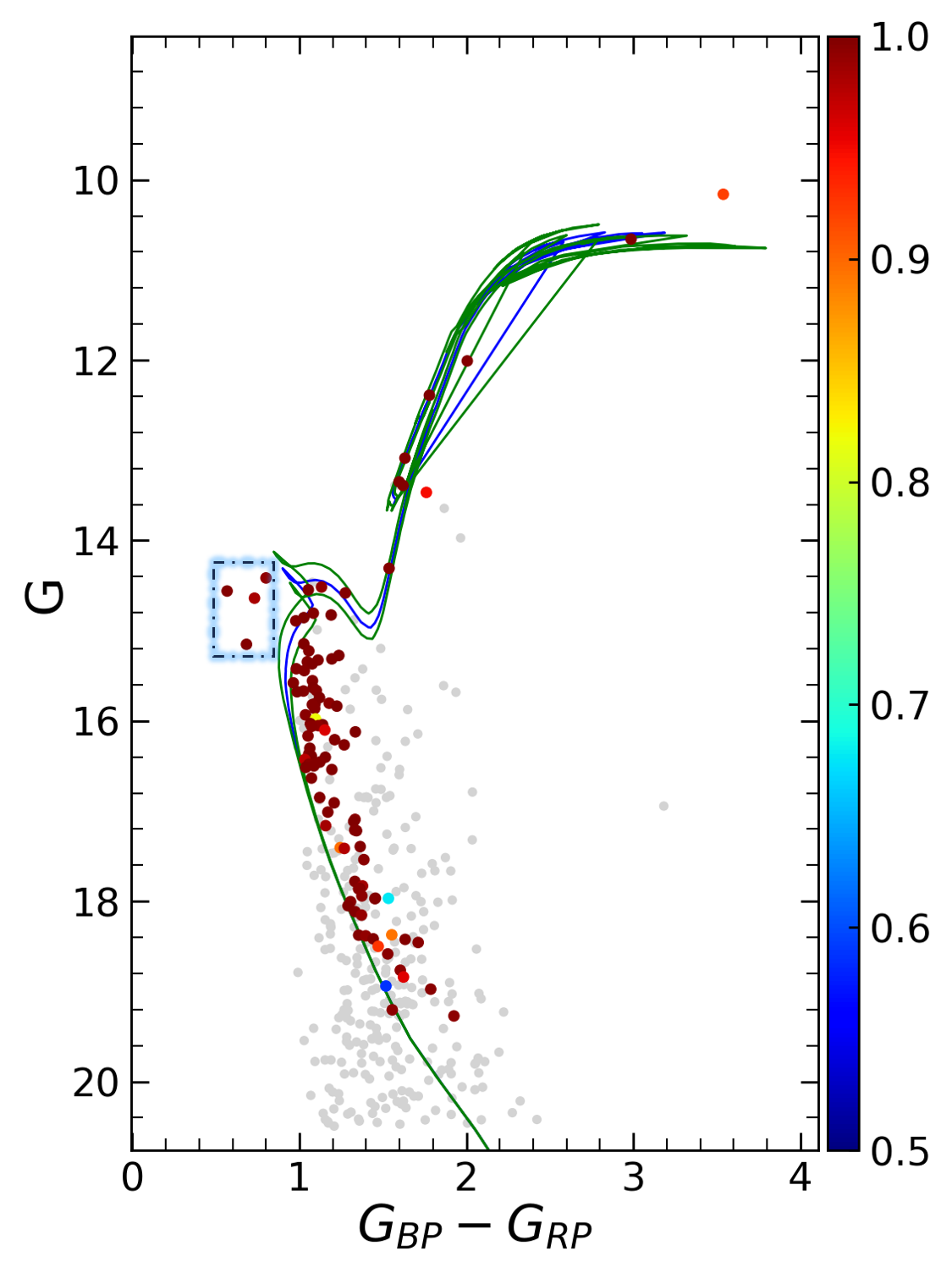}
\caption{Colour-magnitude diagram for the studied cluster Coll 74. Different colour and colourbar scales show the membership probabilities of stars with $P\geq 0.5$. Stars with probabilities $P<0.5$ are demonstrated with filled grey circles. BSSs of the cluster are shown in a blue dashed-lined box. The best solution of fitted isochrones and their errors are inferred as the blue and green lines, respectively. The age of the blue-lined isochrone matches 1800 Myr for the cluster.
\label{fig:figure_age} }
\end {figure}

We plotted $G\times (G_{\rm BP}-G_{\rm RP})$ and superimposed isochrones, scaled to the $z=0.0136$, of different ages ($\log t$=9.20, 9.25 and 9.30 yr) by visual inspection to the most likely cluster main-sequence, turn-off and giant members with probabilities over $P\geq 0.5$ as shown in Figure ~\ref{fig:figure_age}. The best fit supports the isochrone with $\log t$=9.25 yr to the cluster morphology, this isochrone corresponding to $t=1800\pm 200$ Myr. The estimated age is comparable with the values of \citet{Tadross_2001} and \citet{Cantat-Gaudin_2020}. Also, good isochrone fitting result supplies the distance modulus and colour excess of the Coll 74 to be $\mu_{\rm G}=13.052\pm0.088$ mag, corresponding to isochrone distance $d_{\rm iso}=2831\pm118$ pc and $E(G_{\rm BP}-G_{\rm RP})=0.425\pm 0.046$ mag, respectively. We used the relations of \citet{Carraro_2017} to estimate the errors of distance modulus and isochrone distance. Our derived isochrone distance is compatible with most of the studies presented by different researchers (see Table~\ref{tab:literature}) as well as the trigonometric parallax distance $d_{\rm \varpi}=2755\pm 326$ pc estimated in this study. For a more accurate comparison with literature studies, we converted this value to the $U\!BV$-based colour excess $E(B-V)$. We utilised the equation $E(G_{\rm BP}-G_{\rm RP})= 1.41\times E(B-V)$ given by \citet{Sun_2021} and determined the value as $E(B-V)=0.301\pm 0.033$ mag. This result is close to the values given by \citet{Tadross_2001}, \citet{Lada_2002}, \citet{Carraro_2007} and \citet{Hasegawa_2008} within the errors (see Table~\ref{tab:literature}).

We also estimated heliocentric Galactic coordinates $(X, Y, Z)_{\odot}$ of Coll 74. Here, $X$ is the distance from the Galactic center in the Galactic plane ($l=0^{\rm o}$, $b=0^{\rm o}$), $Y$ is the distance in the direction of Galactic rotation ($l=90^{\rm o}$, $b=0^{\rm o}$) and $Z$ is the vertical distance from Galactic plane to the North Galactic Pole ($l=0^{\rm o}$, $b=90^{\rm o}$). Galactocentric coordinates provide a convenient way to describe the positions of celestial objects relative to the Galactic center, Sun, and Galactic plane. By considering isochrone distance, Galactic longitude, and latitude of the cluster, we derived these distances as $(X, Y, Z)_{\odot}=(-2633, -907, -510)$ pc.

\section{Galactic Orbit Study of the Collinder 74}

The Galactic orbits of open clusters are important for understanding how these celestial objects dynamically evolve within the Milky Way \citep{Tasdemir_2023}. We derived the orbits and orbital parameters of Coll 74 with the help of the python based Galactic dynamics library {\sc galpy}\footnote{See also https://galpy.readthedocs.io/en/v1.5.0/} of \citet{Bovy_2015}. This library implements {\sc MWPotential2014} model, which commonly uses a potential model in Galactic dynamics. The {\sc MWPotential2014} model is based on a combination of different components that represent the various structures within the Milky Way, including the bulge, disc, and halo. These components are parameterised to approximate the observed properties of the Galaxy: the bulge is modelled as a power-law density profile as described in \citet{Bovy_2015}, the disc is typically modelled as an exponential disk with a specified scale length and scale height as defined by \citet{Miyamoto_1975} and the halo is often modelled as a spherical or ellipsoidal distribution with a specified density profile as defined by  \citet{Navarro_1996}. We accepted Sun's Galactocentric distance and orbital velocity as $R_{\rm gc}=8$ kpc and $V_{\rm rot}=220$ km s$^{-1}$, respectively \citep{Bovy_2015, Bovy_2012}, as well as Sun's distance from the Galactic plane was adopted as $25 \pm 5$ pc \citep{Juric_2008}. 

The mean radial velocity ($V_{\gamma}$) of the cluster was calculated from available {\it Gaia} DR3 radial velocity measurements of the stars. We considered the most likely member stars with probabilities over $P\geq 0.5$ whose number are 16. We used the equations of \citet{Soubiran_2018} which are based on the weighted average of the radial velocities of the stars. Hence, the mean radial velocity of the Coll 74 was determined as  $V_{\gamma}= 20.55 \pm 0.41$ km s$^{-1}$ which is in good agreement with mean radial velocity findings presented by \citet{Soubiran_2018}, \citet{Dias_2021}, \citet{Tarricq_2021} and \citet{Hunt_2023}. To estimate orbital parameters, we used equatorial coordinates ($\alpha=05^{\rm h} 48^{\rm m} 40^{\rm s}.8$, $\delta= +07^{\rm o} 22^{\rm '} 26^{\rm''}.4$) taken from \citet{Cantat-Gaudin_2020}, mean proper-motion components ($\mu_{\alpha}\cos\delta = 0.960\pm0.005$, $\mu_{\delta}= -1.526\pm0.004$ mas yr$^{-1}$), isochrone distance ($d_{\rm iso}=2831\pm 118$ pc) and the radial velocity ($V_{\gamma}=20.55\pm 0.41$ km s$^{-1}$) calculated in the study (see also Table~\ref{tab:Final_table}) for Coll 74 as input parameters. 

\begin{figure}
\centering
\includegraphics[scale=0.35, angle=0]{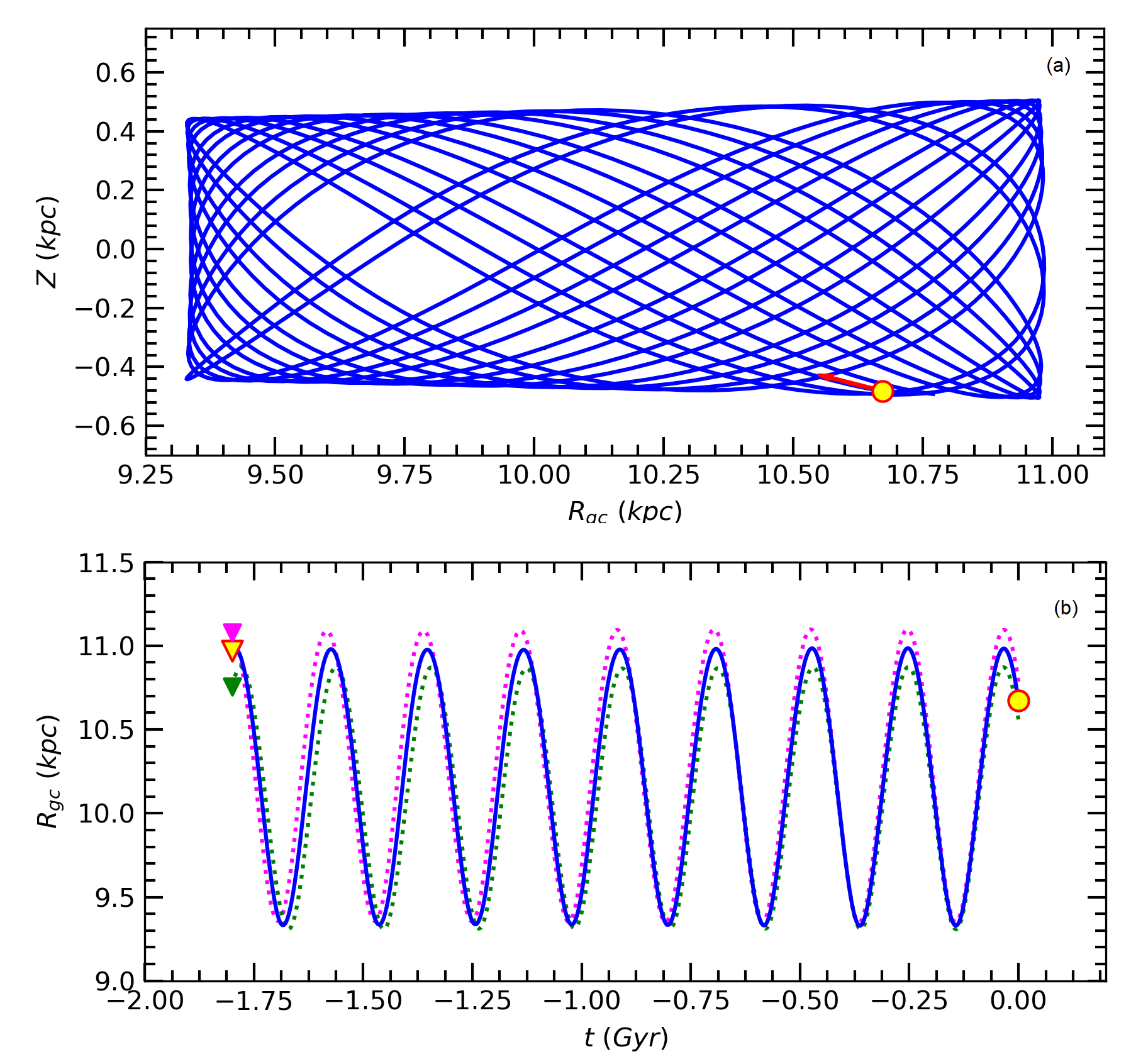}
\small
\caption{The Galactic orbits and birth radii of Coll 74 in the $Z \times R_{\rm gc}$ (a) and $R_{\rm gc} \times t$ (b) planes. The filled yellow circles and triangles show the present-day and birth positions, respectively. The red arrow is the motion vector of Collinder 74. The green and pink dotted lines show the orbit when errors in input parameters are considered, while the green and pink filled triangles represent the birth locations of the open cluster based on the lower and upper error estimates.
\label{fig:galactic_orbits}}
\end {figure}

The orbit integration was applied for the forward with an integration step of 1 Myr up to 2.5 Gyr to estimate the possible current position of Coll 74. The resultant orbit is shown in Figure~\ref{fig:galactic_orbits}a. The figure pictures the path followed by the cluster in $Z \times R_{\rm gc}$ plane, which represents the side view of the orbit. Here, $Z$ and $R_{\rm gc}$ are the distance from the Galactic plane and the Galactic center, respectively. Also, the orbit analyses were carried out for the past epoch across a time equal to $t=1800 \pm 200$ Myr cluster's age. Figure~\ref{fig:galactic_orbits}b shows the cluster's distance variation in time on the $R_{\rm gc} \times t$ plane. The figure also represents the influence of errors in the input parameters on the orbit of Coll 74. Orbit analyses stated that the Coll 74 was formed outside the solar vicinity with a birth radius of $R_{\rm gc}=10.97\pm0.32$ kpc. 

From the orbit integration we derived the following parameters for Coll 74: apogalactic ($R_{\rm a}=10987\pm 112$ pc) and perigalactic ($R_{\rm p}=9337\pm20$ pc) distances, eccentricity ($e=0.081\pm0.004$), maximum vertical distance from Galactic plane ($Z_{\rm max}=506\pm 22$ pc), space velocity components ($U, V, W = -11.43\pm 0.79$, $-29.50\pm 1.08$, $-2.56\pm 0.05$ km s$^{-1}$), and orbital period ($P_{\rm orb}=291\pm2$ Myr). The Local Standard of Rest (LSR) correction was applied to the $(U, V, W)$ components of the Coll 74. To do this, we considered the space velocity component values $(U, V, W)_{\odot}=(8.83\pm 0.24, 14.19\pm 0.34, 6.57\pm0.21$) km s$^{-1}$ of \citet{Coskunoglu_2011}. Hence, LSR corrected space velocity components were found to be $(U, V, W)_{\rm LSR}$ = ($-2.60\pm0.25$, $-15.31\pm1.13$, $4.01\pm0.22$) km s$^{-1}$. Total space velocity was estimated as $S_{\rm LSR}=16.04\pm1.18$ km s$^{-1}$, which is compatible with the velocity value given for thin-disc objects \citep{Leggett_1992}. We interpreted from the perigalactic and apogalactic distances that Coll 74 is completely outside the solar circle (Figure~\ref{fig:galactic_orbits}a). The cluster reaches a maximum distance above the Galactic plane at $Z_{\rm max}=506\pm 22$ pc, shows that Coll 74 belongs to the thin-disc component of the Milky Way \citep{Bilir_2006b, Bilir_2006c, Bilir_2008}.

\section{Luminosity and Mass Functions}
The luminosity function (LF) of an open cluster represents the number of stars at different brightness within the cluster. The LF and mass function (MF) of open clusters are related because the luminosity of a star is generally correlated with its mass. This correlation is also defined as mass-luminosity relation and provides transform LF into the MF \citep{Bisht_2019}.  

\begin{figure}[]
\centering
\includegraphics[scale=1.2, angle=0]{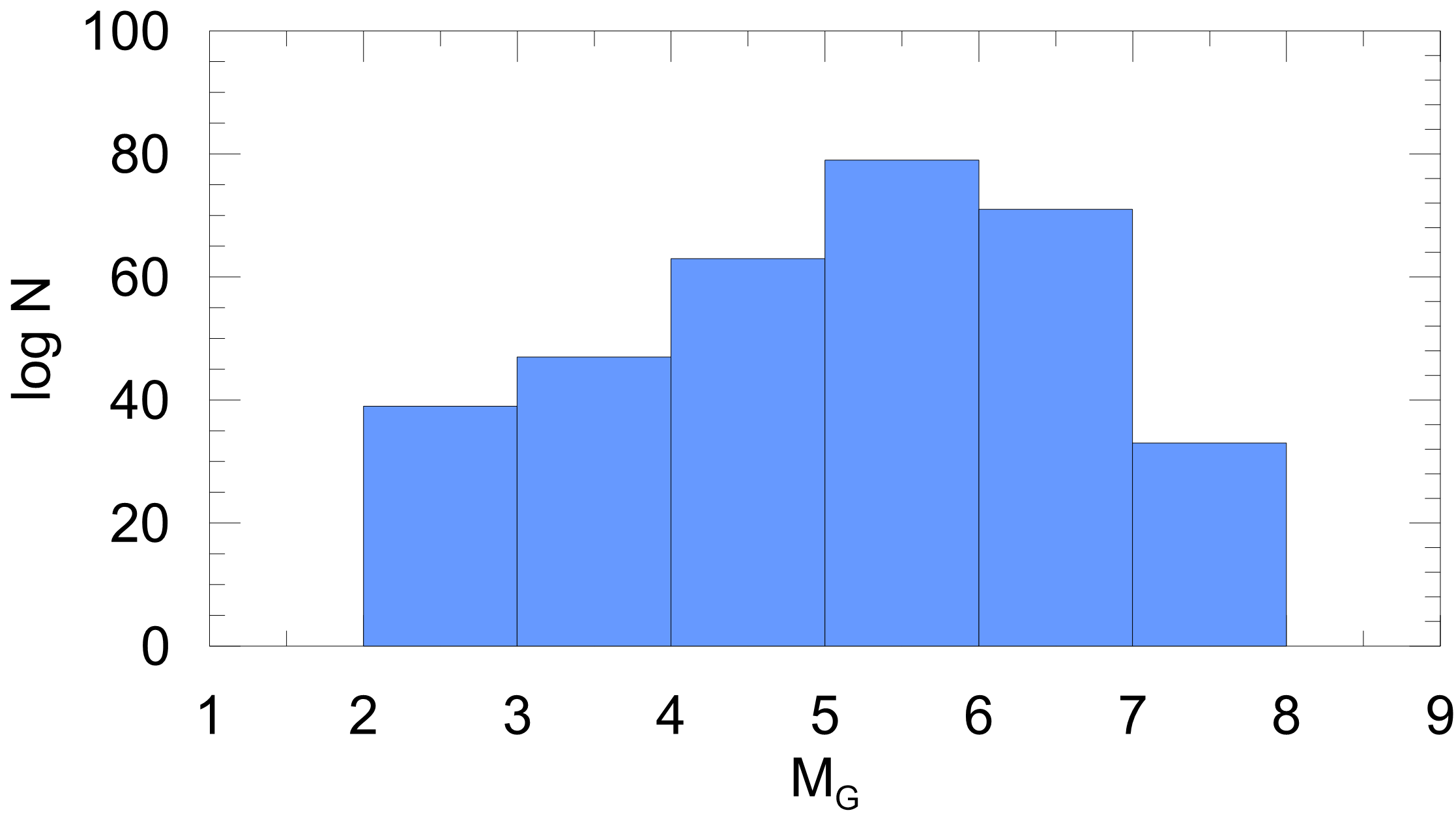}
\caption{\label{fig:luminosity_functions}
The luminosity function of Coll 74.}
\end {figure}

For LF analyses of Coll 74, first, we selected the main-sequence stars with membership probabilities $P>0$ and located inside the limiting radius obtained in the study ($r_{\rm lim}^{\rm obs}=10'$). Hence, we reached 324 stars within the $15\leq G \leq 20.5$ magnitude interval. Then considering distance modulus ($M_{\rm G} = G-5\times \log d+A_{\rm G}$) with apparent magnitude ($G$), isochrone distance ($d_{\rm iso}$) and $G$ band absorption ($A_{\rm G}$) estimated in the study, we transformed apparent $G$ magnitudes into the absolute $M_{\rm G}$ magnitudes. The histogram of LF for the cluster that constructed an interval of 1 mag is shown in Figure~\ref{fig:luminosity_functions}. This figure shows that the number of main-sequence stars rises up to $M_{\rm G}$=6 mag properly, after this limit the counts drop gradually.

\begin{figure}
\centering
\includegraphics[scale=1.4, angle=0]{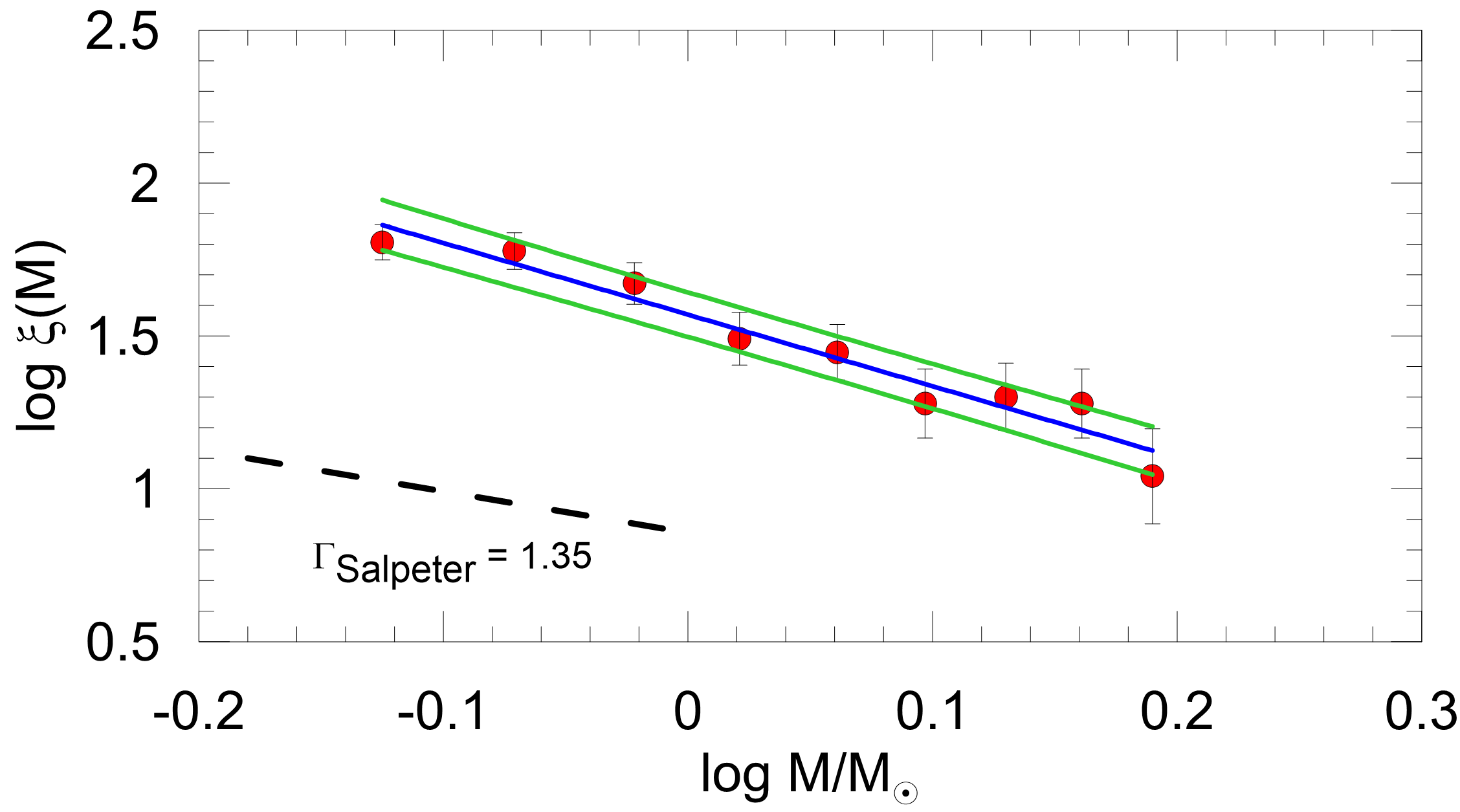}
\caption{\label{fig:mass_functions}
Derived mass function for Coll 74. The blue line represents the MF, whereas the green lines indicate the $\pm1\sigma$ standard deviations. The black dashed line represented \citet{Salpeter_1955}'s slope.}
\end{figure}

For MF estimation, we used {\sc PARSEC} isochrones \citep{Bressan_2012} that scaled to the derived age and adopted metallicity fraction ($z$) for the cluster. From this isochrone, we produced a high-degree polynomial equation between the $G$-band absolute magnitudes and masses. By applying this equation to the selected main-sequence stars ($P>0$), we transformed their absolute magnitudes $M_{\rm G}$ into masses. Hence, we found a mass range of the 324 stars within the $0.65\leq M/ M_{\odot}\leq 1.58$. The MF slope was derived from the following equation:
\begin{eqnarray}
{\rm log(dN/dM)}=-(1+\Gamma)\times \log M + C.
\label{eq:mass_luminosity}
\end{eqnarray}
In the equation $dN$ is the number of stars per unit mass $dM$, $M$ is the central mass, $C$ denotes the constant for the equation, and $\Gamma$ represents the slope of the MF. The estimated MF slope for Coll 74 is $\Gamma=1.34 \pm 0.21$, which is in good agreement with the value of \citet{Salpeter_1955}. The resulting MF is shown in Figure~\ref{fig:mass_functions}.

\begin{figure}[h]
\centering
\includegraphics[scale=.5, angle=0]{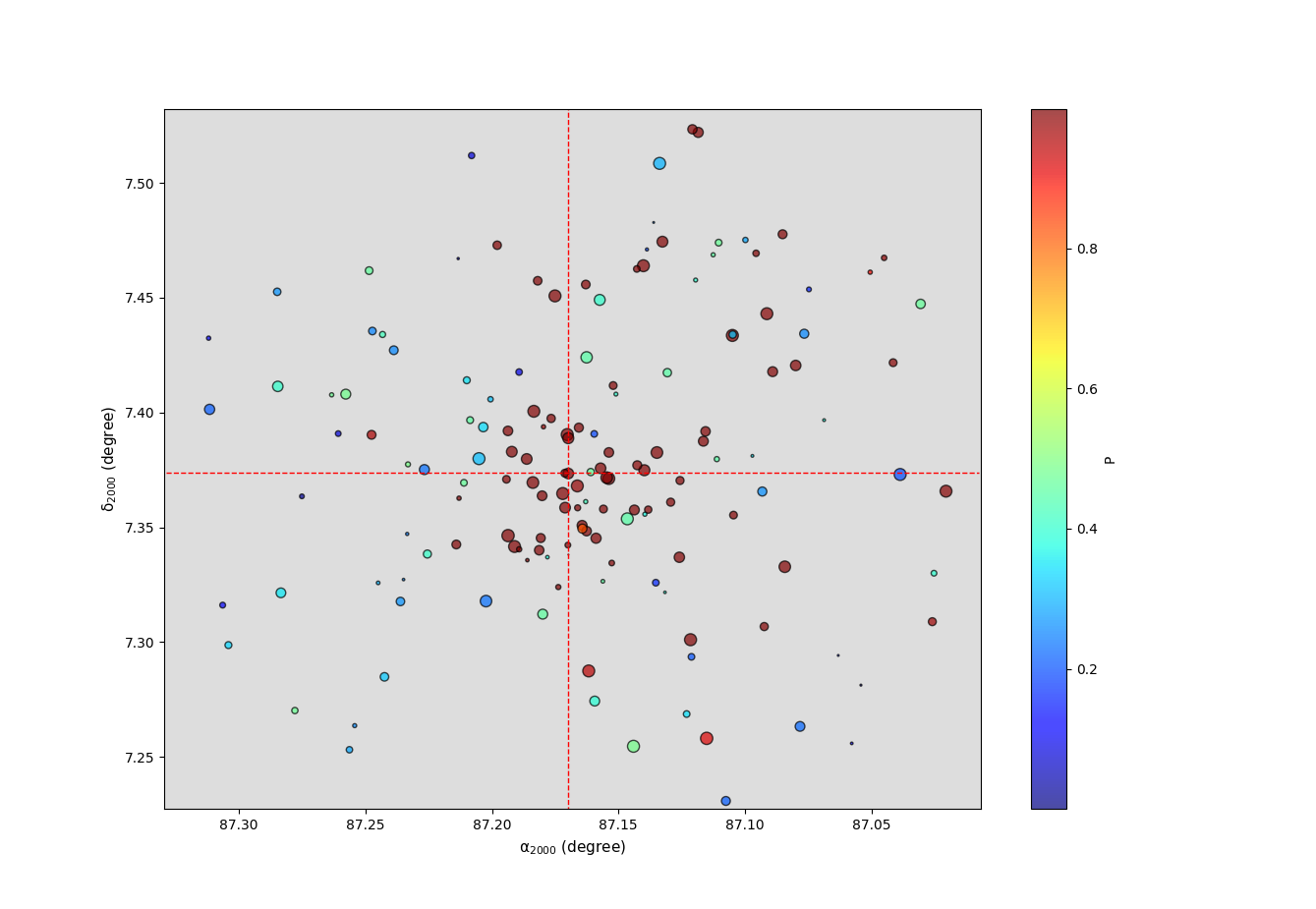}
\caption{\label{fig:mass_distribution}
Mass distribution of the stars in the Coll 74. The radius sizes of the stars indicate the masses, and the different colours show the membership probabilities of the stars. The intersection of the red dashed lines indicates the central position of the cluster in the equatorial coordinate system.}
\end{figure}

The masses of stars in Coll 74 were derived as a function of stars' membership probabilities. The number of stars with probabilities $P>0$ and $P\geq0.5$ was determined as 324 and 102, respectively. Hence, the total mass of the cluster for these probabilities is to be 365$M_{\odot}$ and 132$M_{\odot}$, respectively. We interpreted that the total mass of the cluster estimated from the stars with probabilities $P\geq0.5$ corresponds to about 36$\%$ of the total mass for the stars with all probabilities. To investigate the mass distribution of the stars in Coll 74, 324 stars with $P>0$ were plotted according to their equatorial coordinates and membership probabilities as shown in Figure~\ref{fig:mass_distribution}. It can be interpreted from the figure that the stars with probabilities over 0.8 and massive ones are mostly concentrated central region of the cluster, whereas low-mass stars with probabilities under 0.8 are distributed beyond the cluster center. This case shows that the Coll 74 is a mass segregated open cluster.  

\section{Conclusion}
We performed a detailed {\it Gaia} DR3 data-based study of open cluster Collinder 74. The number of member stars with probabilities over 0.5 were 102. Considering these stars, we calculated structural and fundamental astrophysical parameters, investigated luminosity and mass functions, and estimated the orbit of the cluster. All parameters obtained in the study are listed in Table~\ref{tab:Final_table}. The main results of the study are summarized as follows:

\begin{table}
\renewcommand{\arraystretch}{1.2}
\setlength{\tabcolsep}{14pt}
  \centering
  \caption{~Fundamental parameters of Coll 74.}
  {\normalsize
        \begin{tabular}{lr}
\hline
Parameter & Value\\
\hline
($\alpha,~\delta)_{\rm J2000}$ (Sexagesimal)& 05:48:40.8, $+$07:22:26.4  \\
($l, b)_{\rm J2000}$ (Decimal)              & 199.0189, $-10.3791$     \\    
$f_{0}$ (stars arcmin$^{-2}$)               & $8.42\pm 0.35$           \\
$f_{\rm bg}$ (stars arcmin$^{-2}$)          & $5.45\pm 0.16$           \\
$r_{\rm c}$ (arcmin)                        & $1.38\pm 0.12$           \\
$r_{\rm lim}$ (arcmin)                      & 10                       \\
$r$ (pc)                                    & 8.24                     \\
Cluster members ($P\geq0.5$)                & 102                      \\
$\mu_{\alpha}\cos \delta$ (mas yr$^{-1}$)   & $0.960  \pm 0.005$       \\
$\mu_{\delta}$ (mas yr$^{-1}$)              & $-1.526  \pm 0.004$      \\
$\varpi$ (mas)                              & $0.363 \pm 0.043$        \\
$d_{\varpi}$ (pc)                           & $2755\pm 326$            \\
$E(B-V)$ (mag)                              & $0.301 \pm 0.033$        \\
$E(G_{\rm BP}-G_{\rm RP})$ (mag)            & $0.425\pm 0.046$         \\
$A_{\rm G}$ (mag)                           & $0.792\pm 0.086$         \\
$[{\rm Fe/H}]$ (dex)$^{*}$                  & $-0.052\pm 0.034$        \\
Age (Myr)                                   & $1800\pm 200$            \\
Distance modulus (mag)                      & $13.052  \pm 0.088$      \\
Isochrone distance (pc)                     & $2831 \pm 118$           \\
$(X, Y, Z)_{\odot}$ (pc)                    & ($-2633$, $-907$, $-510$)\\
$R_{\rm gc}$ (kpc)                          & 10.67                    \\
MF slope                                    & $1.34\pm 0.21$           \\
Total mass ($M/M_{\odot})$   ($P>0$)        & 365                      \\
$V_{\gamma}$ (km s$^{-1}$)                  & $20.55 \pm 0.41$         \\
$U_{\rm LSR}$ (km s$^{-1}$)                 & $-2.60 \pm 0.25$         \\
$V_{\rm LSR}$ (kms$^{-1}$)                  & $-15.31 \pm 1.13$        \\
$W_{\rm LSR}$ (kms$^{-1}$)                  & $4.01 \pm0.22$           \\
$S_{_{\rm LSR}}$ (kms$^{-1}$)               & $16.04 \pm1.18$          \\
$R_{\rm a}$ (pc)                            & $10987\pm 112$           \\
$R_{\rm p}$ (pc)                            & $9337 \pm 20$            \\
$z_{\rm max}$ (pc)                          & $506\pm 22$              \\
$e$                                         & $0.081\pm 0.004$         \\
$P_{\rm orb}$ (Myr)                         & $291\pm 2$               \\
Birthplace (kpc)                            & $10.97 \pm 0.32$         \\
\hline
$^{*}$\citet{Zhong_2020}
        \end{tabular}%
    } 
    \label{tab:Final_table}%
\end{table}%

\begin{enumerate}
\item{From the RDP analyses, we determined the limiting radius by visual inspection as $r_{\rm lim}^{\rm obs}=10^{'}$.} 

\item{Considering results of photometric completeness limit, membership probability analyses, and limiting radius, we identified 102 most likely members with probabilities $P\geq0.5$ for Coll 74. These stars were used in the cluster analyses.}

\item{The mean proper-motion components were obtained as ($\mu_{\alpha}\cos \delta, \mu_{\delta})=(0.960 \pm 0.005, -1.526 \pm 0.004$) mas yr$^{-1}$.}

\item{Four most probable BSS members were identified within the limiting radius of the cluster. We concluded that the Coll 74 belongs to a family I according to the radial distribution of its BSSs.}

\item{The metallicity value for the cluster was adopted as $[{\rm Fe/H}]=-0.052 \pm 0.034$ dex which is presented by \citet{Zhong_2020}. We transformed this value into the mass fraction $z=0.0136$ and kept it as a constant parameter for the age and distance modulus estimation.}

\item{By fitting {\sc PARSEC} isochrone \citep{Bressan_2012} to the $G$ versus ($G_{\rm BP}-G_{\rm RP}$) colour-magnitude diagram, we estimated colour excess of the Coll 74 as $E(G_{\rm BP}-G_{\rm RP})=0.425\pm 0.046$} mag, which corresponds to a colour excess in $U\!BV$ system to be $E(B-V)=0.301\pm 0.033$ mag. We estimated this value by using the equation $E(G_{\rm BP}-G_{\rm RP})=1.41\times E(B-V)$ as given by \citet{Sun_2021}.

\item{The isochrone fitting distance of Coll 74 was determined as $d_{\rm iso}=2831\pm 118$ pc. This value is supported by the distance $d_{\varpi}$= $2755\pm 326$ pc that derived from mean trigonometric parallax.}

\item{{\sc PARSEC} isochrone of \citet{Bressan_2012} provides the age of the cluster to be $t=1800\pm 200$ Myr.}

\item{The LF and MF were investigated from the main-sequence stars with probabilities $P>0$. The MF slope was found as $\Gamma=1.34\pm 0.21$ which is in good agreement with the value of \citet{Salpeter_1955}.}

\item{Orbit integration was performed via {\sc MWPotential2014} model. We concluded that Coll 74 orbits in a boxy pattern outside the solar circle, as well as the cluster, is a member of the thin-disc component of the Milky Way. Moreover, the birth radius ($10.97\pm 0.32$ kpc) shows that the forming region of the cluster is outside the solar circle.}

\end{enumerate}

\section*{Acknowledgements}
 This study has been supported in part by the Scientific and Technological Research Council (T\"UB\.ITAK) 122F109. This research has made use of the WEBDA database, operated at the Department of Theoretical Physics and Astrophysics of the Masaryk University. We also made use of NASA's Astrophysics Data System as well as the VizieR and Simbad databases at CDS, Strasbourg, France and data from the European Space Agency (ESA) mission \emph{Gaia}\footnote{https://www.cosmos.esa.int/gaia}, processed by the \emph{Gaia} Data Processing and Analysis Consortium (DPAC)\footnote{https://www.cosmos.esa.int/web/gaia/dpac/consortium}. Funding for DPAC has been provided by national institutions, in particular, the institutions participating in the \emph{Gaia} Multilateral Agreement.



\bibliographystyle{mnras}
\bibliography{refs}

\begin{thebibliography}{}
\makeatletter
\relax
\def\mn@urlcharsother{\let\do\@makeother \do\$\do\&\do\#\do\^\do\_\do\%\do\~}
\def\mn@doi{\begingroup\mn@urlcharsother \@ifnextchar [ {\mn@doi@} {\mn@doi@[]}}
\def\mn@doi@[#1]#2{\def\@tempa{#1}\ifx\@tempa\@empty \href {http://dx.doi.org/#2} {doi:#2}\else \href {http://dx.doi.org/#2} {#1}\fi \endgroup}
\def\mn@eprint#1#2{\mn@eprint@#1:#2::\@nil}
\def\mn@eprint@arXiv#1{\href {http://arxiv.org/abs/#1} {{\tt arXiv:#1}}}
\def\mn@eprint@dblp#1{\href {http://dblp.uni-trier.de/rec/bibtex/#1.xml} {dblp:#1}}
\def\mn@eprint@#1:#2:#3:#4\@nil{\def\@tempa {#1}\def\@tempb {#2}\def\@tempc {#3}\ifx \@tempc \@empty \let \@tempc \@tempb \let \@tempb \@tempa \fi \ifx \@tempb \@empty \def\@tempb {arXiv}\fi \@ifundefined {mn@eprint@\@tempb}{\@tempb:\@tempc}{\expandafter \expandafter \csname mn@eprint@\@tempb\endcsname \expandafter{\@tempc}}}

\bibitem[\protect\citeauthoryear{{Ann}, {Lee}, {Chun}, {Kim}, {Jeon}  \& {Park}}{{Ann} et~al.}{1999}]{Ann_1999}
{Ann} H.~B.,  {Lee} M.~G.,  {Chun} M.~Y.,  {Kim} S.~L.,  {Jeon} Y.~B.,   {Park} B.~G.,  1999, \mn@doi [Journal of Korean Astronomical Society] {10.48550/arXiv.astro-ph/9905103}, 32, 7

\bibitem[\protect\citeauthoryear{{Bilir}, {Karaali}, {Ak}, {Yaz}  \& {Hamzao{\u{g}}lu}}{{Bilir} et~al.}{2006a}]{Bilir_2006b}
{Bilir} S.,  {Karaali} S.,  {Ak} S.,  {Yaz} E.,   {Hamzao{\u{g}}lu} E.,  2006a, \mn@doi [\na] {10.1016/j.newast.2006.10.001}, 12, 234

\bibitem[\protect\citeauthoryear{{Bilir}, {Karaali}  \& {Gilmore}}{{Bilir} et~al.}{2006b}]{Bilir_2006c}
{Bilir} S.,  {Karaali} S.,   {Gilmore} G.,  2006b, \mn@doi [\mnras] {10.1111/j.1365-2966.2006.09891.x}, 366, 1295

\bibitem[\protect\citeauthoryear{{Bilir}, {Cabrera-Lavers}, {Karaali}, {Ak}, {Yaz}  \& {L{\'o}pez-Corredoira}}{{Bilir} et~al.}{2008}]{Bilir_2008}
{Bilir} S.,  {Cabrera-Lavers} A.,  {Karaali} S.,  {Ak} S.,  {Yaz} E.,   {L{\'o}pez-Corredoira} M.,  2008, \mn@doi [\pasa] {10.1071/AS07026}, 25, 69

\bibitem[\protect\citeauthoryear{{Bisht}, {Yadav}, {Ganesh}, {Durgapal}, {Rangwal}  \& {Fynbo}}{{Bisht} et~al.}{2019}]{Bisht_2019}
{Bisht} D.,  {Yadav} R.~K.~S.,  {Ganesh} S.,  {Durgapal} A.~K.,  {Rangwal} G.,   {Fynbo} J.~P.~U.,  2019, \mn@doi [\mnras] {10.1093/mnras/sty2781}, 482, 1471

\bibitem[\protect\citeauthoryear{{Bisht}, {Zhu}, {Yadav}, {Durgapal}  \& {Rangwal}}{{Bisht} et~al.}{2020}]{Bisht_2020}
{Bisht} D.,  {Zhu} Q.,  {Yadav} R.~K.~S.,  {Durgapal} A.,   {Rangwal} G.,  2020, \mn@doi [\mnras] {10.1093/mnras/staa656}, 494, 607

\bibitem[\protect\citeauthoryear{Bovy}{Bovy}{2015}]{Bovy_2015}
Bovy J.,  2015, \mn@doi [\apjs] {10.1088/0067-0049/216/2/29}, 216, 29

\bibitem[\protect\citeauthoryear{Bovy \& Tremaine}{Bovy \& Tremaine}{2012}]{Bovy_2012}
Bovy J.,  Tremaine S.,  2012, \mn@doi [\apj] {10.1088/0004-637X/756/1/89}, 756, 89

\bibitem[\protect\citeauthoryear{Bressan, Marigo, Girardi, Salasnich, Dal~Cero, Rubele  \& Nanni}{Bressan et~al.}{2012}]{Bressan_2012}
Bressan A.,  Marigo P.,  Girardi L.,  Salasnich B.,  Dal~Cero C.,  Rubele S.,   Nanni A.,  2012, \mn@doi [\mnras] {10.1111/j.1365-2966.2012.21948.x}, 427, 127

\bibitem[\protect\citeauthoryear{Cantat-Gaudin et~al.,}{Cantat-Gaudin et~al.}{2018}]{Cantat-Gaudin_2018}
Cantat-Gaudin T.,  et~al., 2018, \mn@doi [\aap] {10.1051/0004-6361/201833476}, 618, A93

\bibitem[\protect\citeauthoryear{Cantat-Gaudin et~al.,}{Cantat-Gaudin et~al.}{2020}]{Cantat-Gaudin_2020}
Cantat-Gaudin T.,  et~al., 2020, \mn@doi [\aap] {10.1051/0004-6361/202038192}, 640, A1

\bibitem[\protect\citeauthoryear{{Carraro} \& {Costa}}{{Carraro} \& {Costa}}{2007}]{Carraro_2007}
{Carraro} G.,  {Costa} E.,  2007, \mn@doi [\aap] {10.1051/0004-6361:20066350}, \href {https://ui.adsabs.harvard.edu/abs/2007A&A...464..573C} {464, 573}

\bibitem[\protect\citeauthoryear{{Carraro}, {Sales Silva}, {Moni Bidin}  \& {Vazquez}}{{Carraro} et~al.}{2017}]{Carraro_2017}
{Carraro} G.,  {Sales Silva} J.~V.,  {Moni Bidin} C.,   {Vazquez} R.~A.,  2017, \mn@doi [\aj] {10.3847/1538-3881/153/3/99}, 153, 99

\bibitem[\protect\citeauthoryear{Co{\c{s}}kuno{\v{g}}lu et~al.,}{Co{\c{s}}kuno{\v{g}}lu et~al.}{2011}]{Coskunoglu_2011}
Co{\c{s}}kuno{\v{g}}lu B.,  et~al., 2011, \mn@doi [\mnras] {10.1111/j.1365-2966.2010.17983.x}, 412, 1237

\bibitem[\protect\citeauthoryear{{Dias}, {Assafin}, {Fl{\'o}rio}, {Alessi}  \& {L{\'\i}bero}}{{Dias} et~al.}{2006}]{Dias_2006}
{Dias} W.~S.,  {Assafin} M.,  {Fl{\'o}rio} V.,  {Alessi} B.~S.,   {L{\'\i}bero} V.,  2006, \mn@doi [\aap] {10.1051/0004-6361:20052741}, 446, 949

\bibitem[\protect\citeauthoryear{Dias, Monteiro, Caetano, L{\'e}pine, Assafin  \& Oliveira}{Dias et~al.}{2014}]{Dias_2014}
Dias W.~S.,  Monteiro H.,  Caetano T.~C.,  L{\'e}pine J.~R.~D.,  Assafin M.,   Oliveira A.~F.,  2014, \mn@doi [\aap] {10.1051/0004-6361/201323226}, 564, A79

\bibitem[\protect\citeauthoryear{{Dias}, {Monteiro}  \& {Assafin}}{{Dias} et~al.}{2018}]{Dias_2018}
{Dias} W.~S.,  {Monteiro} H.,   {Assafin} M.,  2018, \mn@doi [\mnras] {10.1093/mnras/sty1456}, 478, 5184

\bibitem[\protect\citeauthoryear{Dias, Monteir, Moitinho, L{\'e}pine, Carraro, Paunzen, Alessi  \& Villela}{Dias et~al.}{2021}]{Dias_2021}
Dias W.~S.,  Monteir H.,  Moitinho A.,  L{\'e}pine J.~R.~D.,  Carraro G.,  Paunzen E.,  Alessi B.,   Villela L.,  2021, \mn@doi [\mnras] {10.1093/mnras/stab770}, 504, 356

\bibitem[\protect\citeauthoryear{{Ferraro} et~al.,}{{Ferraro} et~al.}{2012}]{Ferraro_2012}
{Ferraro} F.~R.,  et~al., 2012, \mn@doi [\nat] {10.1038/nature11686}, 492, 393

\bibitem[\protect\citeauthoryear{{Gaia Collaboration} et~al.,}{{Gaia Collaboration} et~al.}{2016}]{Gaia_DR1}
{Gaia Collaboration} et~al., 2016, \mn@doi [\aap] {10.1051/0004-6361/201629272}, 595, A1

\bibitem[\protect\citeauthoryear{{Gaia Collaboration} et~al.,}{{Gaia Collaboration} et~al.}{2018}]{Gaia_DR2}
{Gaia Collaboration} et~al., 2018, \mn@doi [\aap] {10.1051/0004-6361/201833051}, 616, A1

\bibitem[\protect\citeauthoryear{{Gaia Collaboration} et~al.,}{{Gaia Collaboration} et~al.}{2023}]{Gaia_DR3}
{Gaia Collaboration} et~al., 2023, \mn@doi [\aap] {10.1051/0004-6361/202243940}, 674, A1

\bibitem[\protect\citeauthoryear{{Hasegawa}, {Sakamoto}  \& {Malasan}}{{Hasegawa} et~al.}{2008}]{Hasegawa_2008}
{Hasegawa} T.,  {Sakamoto} T.,   {Malasan} H.~L.,  2008, \mn@doi [\pasj] {10.1093/pasj/60.6.1267}, 60, 1267

\bibitem[\protect\citeauthoryear{{Hills} \& {Day}}{{Hills} \& {Day}}{1976}]{Hills_1976}
{Hills} J.~G.,  {Day} C.~A.,  1976, \aplett, \href {https://ui.adsabs.harvard.edu/abs/1976ApL....17...87H} {17, 87}

\bibitem[\protect\citeauthoryear{{Hunt} \& {Reffert}}{{Hunt} \& {Reffert}}{2023}]{Hunt_2023}
{Hunt} E.~L.,  {Reffert} S.,  2023, \mn@doi [\aap] {10.1051/0004-6361/202346285}, \href {https://ui.adsabs.harvard.edu/abs/2023A&A...673A.114H} {673, A114}

\bibitem[\protect\citeauthoryear{{Jadhav} \& {Subramaniam}}{{Jadhav} \& {Subramaniam}}{2021}]{Jadhav_2021}
{Jadhav} V.~V.,  {Subramaniam} A.,  2021, \mn@doi [\mnras] {10.1093/mnras/stab2264}, 507, 1699

\bibitem[\protect\citeauthoryear{{Juri{\'c}} et~al.,}{{Juri{\'c}} et~al.}{2008}]{Juric_2008}
{Juri{\'c}} M.,  et~al., 2008, \mn@doi [\apj] {10.1086/523619}, 673, 864

\bibitem[\protect\citeauthoryear{Kharchenko, Piskunov, Schilbach, R{\"o}ser  \& Scholz}{Kharchenko et~al.}{2013}]{Kharchenko_2013}
Kharchenko N.~V.,  Piskunov A.~E.,  Schilbach E.,  R{\"o}ser S.,   Scholz R.~D.,  2013, \mn@doi [\aap] {10.1051/0004-6361/201322302}, 558, A53

\bibitem[\protect\citeauthoryear{{Kim}, {Kyeong}, {Park}, {Han}, {Lee}, {Moon}, {Lee}  \& {Kim}}{{Kim} et~al.}{2017}]{Kim_2017}
{Kim} S.~C.,  {Kyeong} J.,  {Park} H.~S.,  {Han} I.,  {Lee} J.~H.,  {Moon} D.-S.,  {Lee} Y.,   {Kim} S.,  2017, \mn@doi [Journal of Korean Astronomical Society] {10.5303/JKAS.2017.50.3.79}, 50, 79

\bibitem[\protect\citeauthoryear{King}{King}{1962}]{King_1962}
King I.,  1962, \mn@doi [\aj] {10.1086/108756}, 67, 471

\bibitem[\protect\citeauthoryear{Krone-Martins \& Moitinho}{Krone-Martins \& Moitinho}{2014}]{Krone-Martins_2014}
Krone-Martins A.,  Moitinho A.,  2014, \mn@doi [\aap] {10.1051/0004-6361/201321143}, 561, A57

\bibitem[\protect\citeauthoryear{Lada \& Lada}{Lada \& Lada}{2003}]{Lada_2003}
Lada C.~J.,  Lada E.~A.,  2003, \mn@doi [\araa] {10.1146/annurev.astro.41.011802.094844}, 41, 57

\bibitem[\protect\citeauthoryear{{Lata}, {Pandey}, {Sagar}  \& {Mohan}}{{Lata} et~al.}{2002}]{Lada_2002}
{Lata} S.,  {Pandey} A.~K.,  {Sagar} R.,   {Mohan} V.,  2002, \mn@doi [\aap] {10.1051/0004-6361:20020450}, 388, 158

\bibitem[\protect\citeauthoryear{Leggett}{Leggett}{1992}]{Leggett_1992}
Leggett S.~K.,  1992, \mn@doi [\apjs] {10.1086/191720}, 82, 351

\bibitem[\protect\citeauthoryear{Liu \& Pang}{Liu \& Pang}{2019}]{Liu_2019}
Liu L.,  Pang X.,  2019, \mn@doi [\apjs] {10.3847/1538-4365/ab530a}, 245, 32

\bibitem[\protect\citeauthoryear{Loktin \& Popova}{Loktin \& Popova}{2017}]{Loktin_2017}
Loktin A.~V.,  Popova M.~E.,  2017, \mn@doi [Astrophysical Bulletin] {10.1134/S1990341317030154}, 72, 257

\bibitem[\protect\citeauthoryear{{Marsakov}, {Gozha}, {Koval'}  \& {Shpigel'}}{{Marsakov} et~al.}{2016}]{Marsakov_2016}
{Marsakov} V.~A.,  {Gozha} M.~L.,  {Koval'} V.~V.,   {Shpigel'} L.~V.,  2016, \mn@doi [Astronomy Reports] {10.1134/S1063772915120033}, 60, 43

\bibitem[\protect\citeauthoryear{Miyamoto \& Nagai}{Miyamoto \& Nagai}{1975}]{Miyamoto_1975}
Miyamoto M.,  Nagai R.,  1975, \pasj, 27, 533

\bibitem[\protect\citeauthoryear{{Navarro}, {Frenk}  \& {White}}{{Navarro} et~al.}{1996}]{Navarro_1996}
{Navarro} J.~F.,  {Frenk} C.~S.,   {White} S. D.~M.,  1996, \apj, 462, 563

\bibitem[\protect\citeauthoryear{{Rain}, {Ahumada}  \& {Carraro}}{{Rain} et~al.}{2021}]{Rain_2021}
{Rain} M.~J.,  {Ahumada} J.~A.,   {Carraro} G.,  2021, \mn@doi [\aap] {10.1051/0004-6361/202040072}, 650, A67

\bibitem[\protect\citeauthoryear{{Salpeter}}{{Salpeter}}{1955}]{Salpeter_1955}
{Salpeter} E.~E.,  1955, \mn@doi [\apj] {10.1086/145971}, 121, 161

\bibitem[\protect\citeauthoryear{{Sandage}}{{Sandage}}{1953}]{Sandage_1953}
{Sandage} A.~R.,  1953, \mn@doi [\aj] {10.1086/106822}, \href {https://ui.adsabs.harvard.edu/abs/1953AJ.....58...61S} {58, 61}

\bibitem[\protect\citeauthoryear{{Sariya} et~al.,}{{Sariya} et~al.}{2021}]{Sariya_2021}
{Sariya} D.~P.,  et~al., 2021, \mn@doi [\aj] {10.3847/1538-3881/abd31d}, 161, 101

\bibitem[\protect\citeauthoryear{Soubiran et~al.,}{Soubiran et~al.}{2018}]{Soubiran_2018}
Soubiran C.,  et~al., 2018, \mn@doi [\aap] {10.1051/0004-6361/201834020}, 619, A155

\bibitem[\protect\citeauthoryear{Sun, Jiang, Yuan  \& Li}{Sun et~al.}{2021}]{Sun_2021}
Sun M.,  Jiang B.,  Yuan H.,   Li J.,  2021, \mn@doi [\apjs] {10.3847/1538-4365/abf929}, 254, 38

\bibitem[\protect\citeauthoryear{{Ta\c sdemir} \& {Yontan}}{{Ta\c sdemir} \& {Yontan}}{2023}]{Tasdemir_2023}
{Ta\c sdemir} S.,  {Yontan} T.,  2023, \mn@doi [Physics and Astronomy Reports] {10.26650/PAR.2023.00001}, 1, 1

\bibitem[\protect\citeauthoryear{{Tadross}}{{Tadross}}{2001}]{Tadross_2001}
{Tadross} A.~L.,  2001, \mn@doi [\na] {10.1016/S1384-1076(01)00061-6}, 6, 293

\bibitem[\protect\citeauthoryear{Tarricq et~al.,}{Tarricq et~al.}{2021}]{Tarricq_2021}
Tarricq Y.,  et~al., 2021, \mn@doi [\aap] {10.1051/0004-6361/202039388}, 647, A19

\bibitem[\protect\citeauthoryear{{Zacharias}, {Finch}, {Girard}, {Henden}, {Bartlett}, {Monet}  \& {Zacharias}}{{Zacharias} et~al.}{2013}]{Zacharias_2013}
{Zacharias} N.,  {Finch} C.~T.,  {Girard} T.~M.,  {Henden} A.,  {Bartlett} J.~L.,  {Monet} D.~G.,   {Zacharias} M.~I.,  2013, \mn@doi [\aj] {10.1088/0004-6256/145/2/44}, 145, 44

\bibitem[\protect\citeauthoryear{{Zhong}, {Chen}, {Wu}, {Li}, {Bai}  \& {Hou}}{{Zhong} et~al.}{2020}]{Zhong_2020}
{Zhong} J.,  {Chen} L.,  {Wu} D.,  {Li} L.,  {Bai} L.,   {Hou} J.,  2020, \mn@doi [\aap] {10.1051/0004-6361/201937131}, 640, A127

\bibitem[\protect\citeauthoryear{{Zinn} \& {Dahn}}{{Zinn} \& {Dahn}}{1976}]{Zinn_1976}
{Zinn} R.,  {Dahn} C.~C.,  1976, \mn@doi [\aj] {10.1086/111916}, 81, 527

\makeatother
\end{thebibliography}

\bsp	
\label{lastpage}
\end{document}